\documentclass[preprint2]{aastex}
\usepackage{graphicx}
\usepackage{txfonts}
\usepackage{natbib}

\usepackage{lineno}
\usepackage{bm}

\begin{document}

\title{Discovery of Kolmogorov-like turbulence in the inner shell of SNR HB~9}





\author{
L\sc{i} X\sc{iao},\altaffilmark{1,2,3}
M\sc{ing} Z\sc{hu},\altaffilmark{1,2,3}
X\sc{iao}-H\sc{ui} S\sc{un},\altaffilmark{4}
W\sc{olfgang} R\sc{eich},\altaffilmark{5}
P\sc{atricia} R\sc{eich}\altaffilmark{5}     
}

\altaffiltext{1}{National Astronomical Observatories, Chinese Academy of
            Sciences, Jia-20, Datun Road, 100012 Beijing, China; Email: xl@nao.cas.cn}
\altaffiltext{2}{Key Laboratory of FAST, NAOC, Chinese Academy of Science, 100012 Beijing, China}
\altaffiltext{3}{Guizhou Radio Astronomical Observatory, Guizhou University, 100012 Guiyang, China}
\altaffiltext{4}{School of Physics and Astronomy, Yunnan University, 650091 Kunming, China }
\altaffiltext{5}{Max-Planck-Institut f\"{u}r Radioastronomie, Auf dem H\"ugel 69, 53121 Bonn, Germany}


\begin{abstract}
We investigated the scaling relation of the magnetohydrodynamic turbulence in the inner and outer shells of the supernova remnant HB~9 using multi‑tracers.
For the eastern inner shell section, one-dimensional structure function and power spectrum analyses of the radio total intensity, H$_{\alpha}$ emission, and 
polarized intensity consistently show a Kolmogorov‑like scaling $l^{2/3}$ over the inertial range. 
The two-dimensional power spectrum for the entire inner region follows $k^{-8/3}$ of all tracers, including the polarization gradient~$\nabla P$.~
It indicates that fully developed, trans‑Alfvénic Kolmogorov turbulence is present in the inner shell of HB 9.
We estimated the strength of random magnetic field therein from the low fractional polarization ($\sim$5\%), and obtained a upper limit of $\delta B \sim 18~\mu$G.
The turbulence in the inner shock shell is most likely driven by the Richtmyer-Meshkov instability.
\end{abstract}


\maketitle

\section{Introduction}
Supernova remnants (SNRs) are the primary accelerator of Galactic cosmic-rays (CRs), and can accelerate particles to several hundred MeV via the diffusive shock acceleration (DSA) mechanism~\citep{bo78}. 
In this scenario, magnetic turbulence scatters charged particles repeatedly across the shock front and confines them in the acceleration region. 
The energy spectrum of the turbulence directly determines the particle diffusion coefficient and the acceleration efficiency. 
Therefore, measuring the magnetic energy spectrum within SNRs is crucial for understanding the particle acceleration process.

Theoretically, ~\citet{gs95} established that Kolmogorov-type scaling ($E(k) \propto k^{-5/3}$) holds in the direction perpendicular to the local magnetic field, 
for incompressible magnetohydrodynamic (MHD) turbulence, while the spectrum becomes steeper in the parallel direction.
For compressible MHD turbulence, which is more relevant for the interstellar medium and SNR environments,
numerical simulations have greatly improved our understanding of scaling laws, anisotropy, and compressibility~\citep{cl02,b19}.  

Synchrotron radiation emitted by relativistic electrons motions in turbulent magnetic fields provides information on the magnetic field.
Based on the modern understanding of MHD turbulence,~\citet{lp12,lp16} developed a comprehensive statistical framework to extract the characteristics of underlying magnetic turbulence
via structure function (SF) and power spectrum (PS) analyses of fluctuations in synchrotron total intensity, polarized intensity, and rotation measure ($RM$). 
These methods were widely applied to both MHD simulations of diffuse emission and SNRs. 
More recently,~\citet{zl25} proposed the gradient of polarized intensity $\nabla P$~\citep{ghb11} as a robust tracer of the inertial-range scaling index,
with the advantage of being less affected by Faraday depolarization. However, its application to SNR turbulence diagnostics remains unexplored.

Observationally, Kolmogorov-type scaling has been discovered in several young SNRs, with a break identified from auto-correlation and power spectrum analyses of
radio synchrotron total intensity (e.g., Kepler:~\citet{sbr19}; Cas A:~\citet{sbc21}). \citet{sal18} analyzed magnetic energy spectral variations at
different distances from the shock front of Tycho's SNR and explored the roles of Richtmyer–Meshkov instability and cosmic-ray-driven instabilities in driving turbulence and magnetic field amplification. 
Recently, \citet{pk25} confirmed that Kolmogorov-type turbulence has developed in the 3D volume of Tycho's SNR from both radio and X-ray brightness maps. 
For middle-aged SNRs,~\citet{ssa23} reported a relatively shallow spectrum derived from SF of $RMs$ for SNRs G46.8–0.3 ($\sim$10000 yrs) and G39.2–0.3 ($\sim$5000 yrs) in the HI/OH/Recombination line(THOR) survey. The turbulent properties within more SNRs remain to be explored.

HB 9 (G160.9+2.6) is a middle-aged (4000$-$7000~yrs) SNR~\citep{lrg20}, evolving in a relatively low-density cavity environment with a double spherical shell structure~\citep{sey19}.
It has been observed in multi-wavelength observations, including polarized observations at 1.4, 2.7 and 5~GHz~\citep{xzs25}, and
the derived $RM_{2.7/5}$ distribution reveals a large $RM$ in the inner region,
resembling those seen in young SNRs. Its large angular diameter ($\sim 2^\circ$) provides sufficient angular resolution for SF and PS studies using the
Five-hundred-meter Aperture Spherical radio Telescope (FAST)~\citep{jyg19}, enabling investigations of turbulence properties in the shell regions.

In this paper, we present the first comprehensive multi-probe statistical analysis of turbulence in the inner shell of SNR HB 9 using SF and PS techniques applied to total intensity ($I$), 
optical H$\alpha$ emission, polarized intensity ($PI$), and polarization gradient ($\nabla P$).
The paper is organized as follows. Section 2 describes the observational data sets. Section 3 presents the methodology of SF and PS calculation.  
Section 4 reports our results for these observables. Section 5 discusses the physical implications. Conclusions are summarized in Sect. 5.

\section{Data}
\subsection{Radio total-intensity and polarization maps}
To probe the magnetic energy spectrum, we analyzed the FAST 1.4-GHz total intensity data from~\citet{xzs25}, convolved to a resolution of 6$\arcmin$ (shown in Fig.~\ref{images}). 
The pixel size is 1.32$\arcmin$. The image noise level is $\sigma =0.5$~mK. Adopting a distance of 540~pc to HB~9~\citep{zjl20}, 
the pixel size of 1.32$\arcmin$ corresponds to a physical separation of $\sim 0.21$~pc. 
The radio total intensity is sensitive to the perpendicular component of the magnetic field, following $I\propto nB_{\perp}^{1+\alpha}$, where
$n$ is an uniform distribution of the CR electrons density, and the radio spectral index $\alpha$ is 0.5 for the inner shell.

We also use the Urumqi 5-GHz polarization map (Fig.~\ref{images}) to trace fluctuations in $B_{\perp}$~\citep{ghr11}. This data suffers little Faraday rotation,
and therefore can be regarded as intrinsic polarization emission. Its angular resolution is 9.75$\arcmin$.
We present the polarization-fraction map in Fig.~\ref{images}, to reveal the low polarization fraction ($\sim$5\%) in the eastern inner shell.


\subsection{H$_{\alpha}$ intensity} 
To probe the perturbations in electron density $n_{e}$, we used data from the Northern Sky Narrowband Survey~\citep{z25}, which is flux calibrated
with an angular resolution of 9$\arcsec$. 
The map was smoothed to 4$\arcmin$ and resampled to match the pixel size of FAST, and is displayed in Fig.~\ref{images}.
Neglecting absorption, the H$_{\alpha}$ line intensity from neutral hydrogen recombination transitions can be used to derive the emission measure
($EM= \int n^{2}_{e}ds$)~\citep{hrt98} via
%
\begin{equation}\label{Dr}
(\frac{EM}{cm^{-6}pc})= 2.75(\frac{T_{e}}{10^{4}~K})^{0.9}(\frac{I_{H_{\alpha}}}{R})  
\end{equation}

\subsection{Polarization-intensity gradient}
\begin{figure*}[!hbt]
\begin{center}
\includegraphics[angle=0,height=0.42\textwidth]{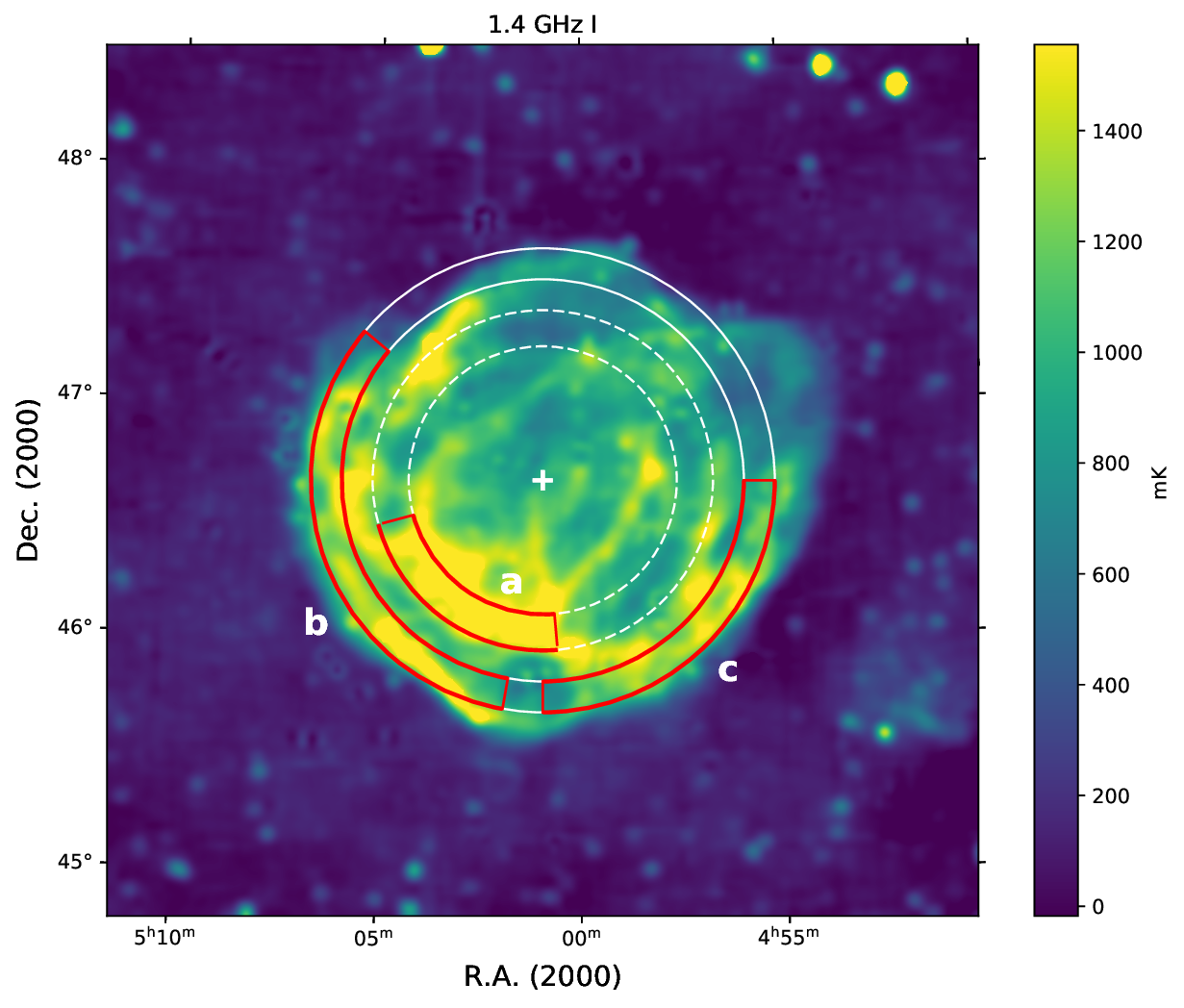}
\includegraphics[angle=0,height=0.42\textwidth]{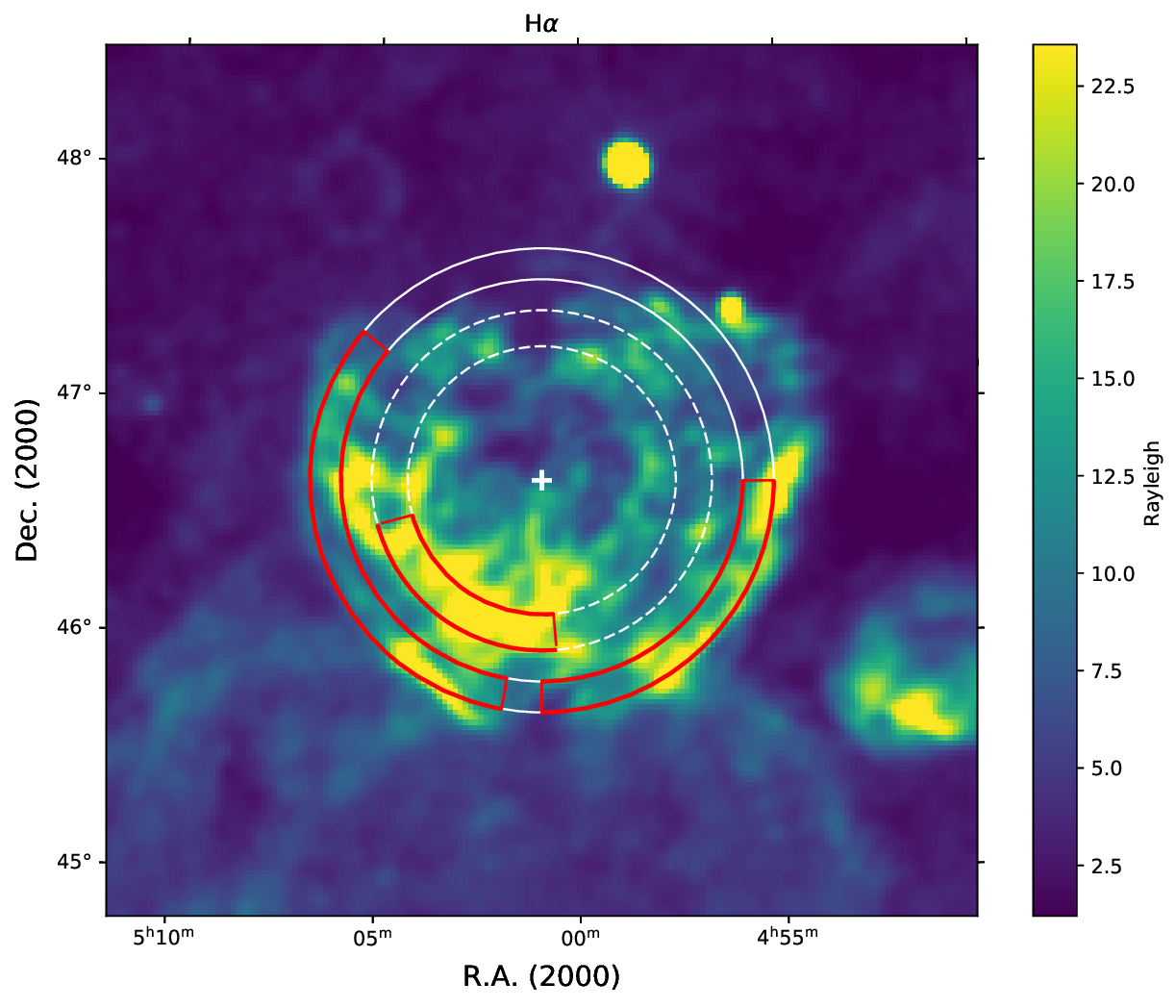}
\includegraphics[angle=0,height=0.42\textwidth]{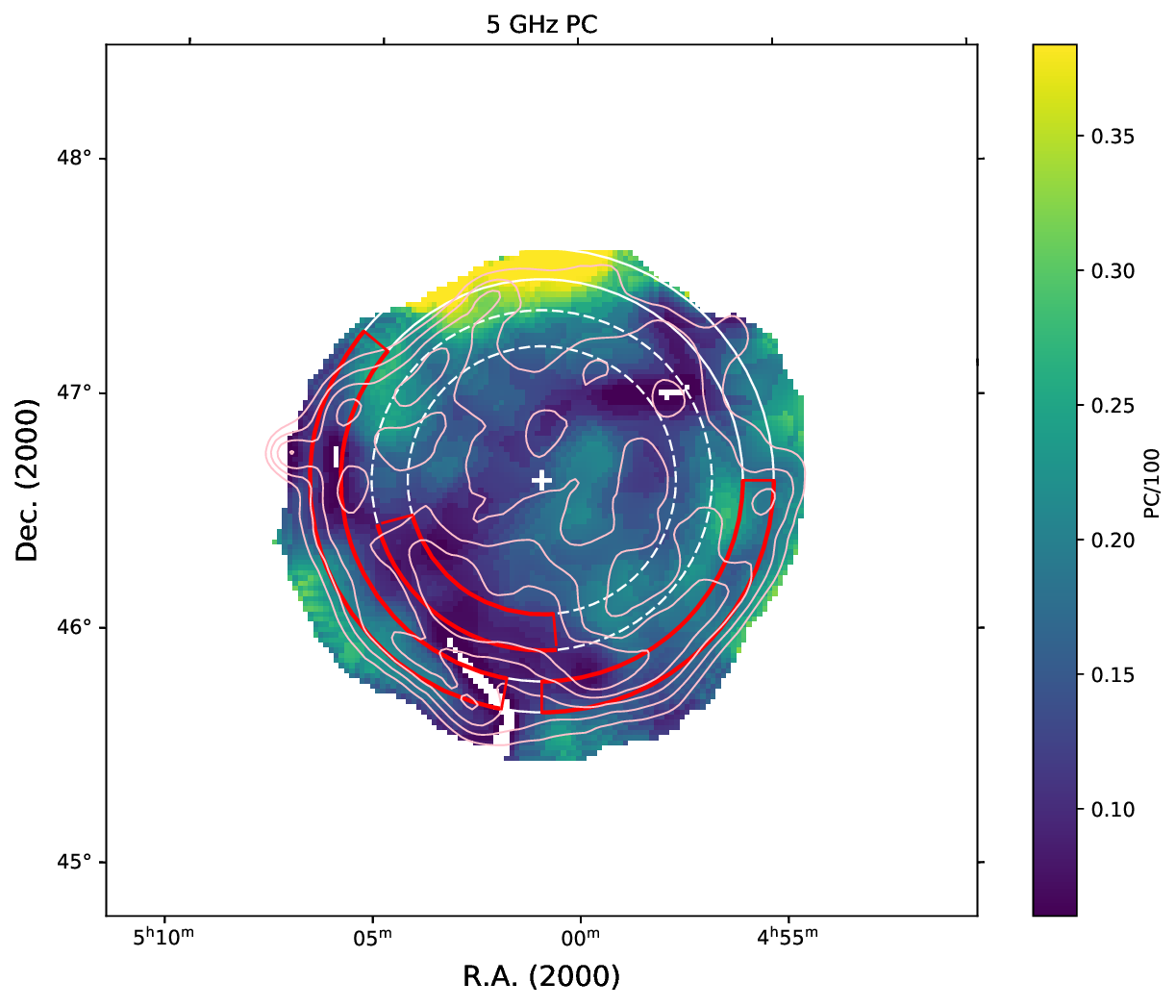}
\includegraphics[angle=0,height=0.42\textwidth]{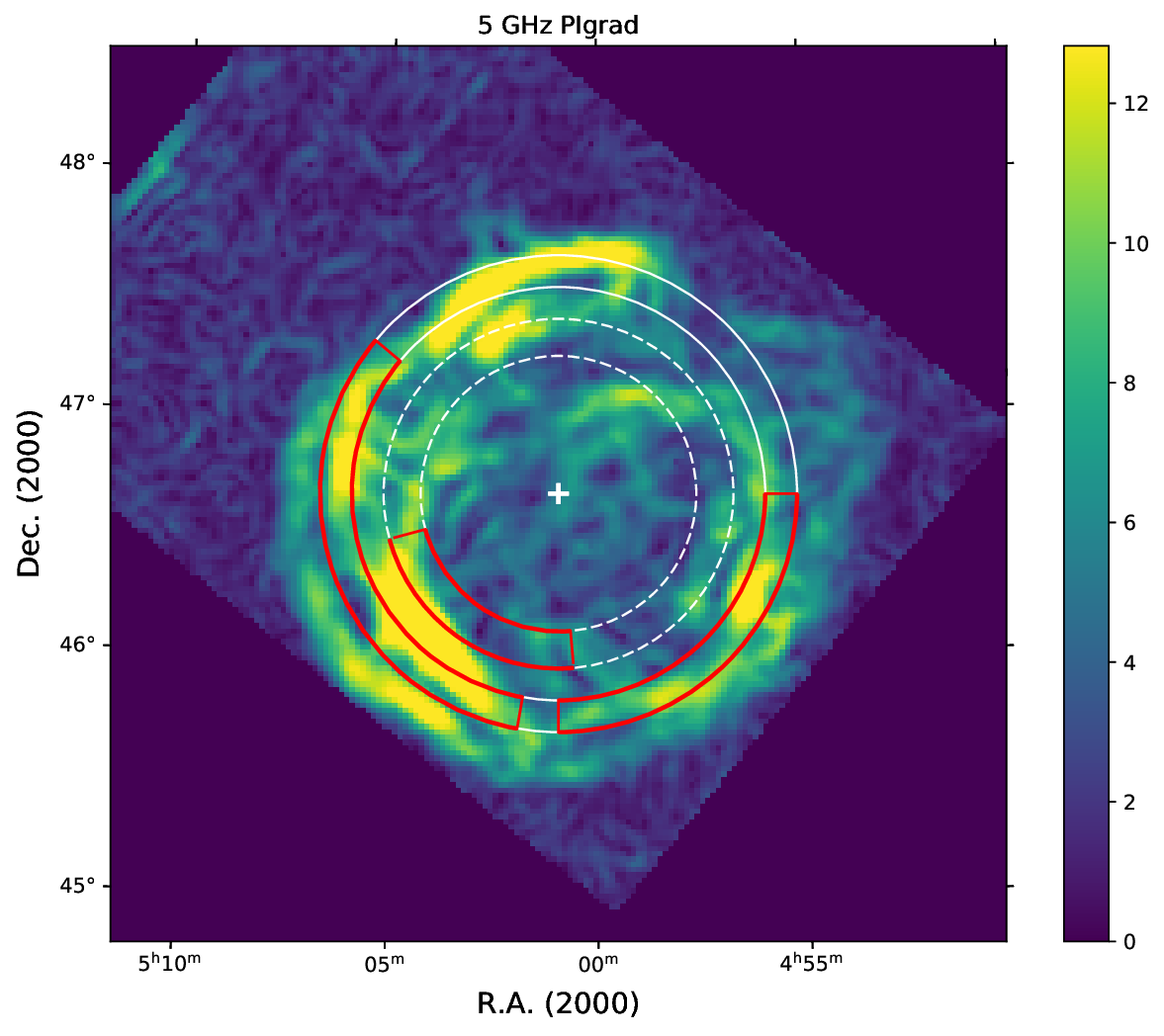}
\caption{$Upper~panel$: The FAST 1.4-GHz total-intensity map of HB~9 and the H$\alpha$ image with an angular resolution of 6$\arcmin$.
$Lower~panel$: The Urumqi 5-GHz polarization-percentage map and the gradient of polarization intensity of HB~9. 
Contours show the total intensities at 5-GHz at 30, 40, 50~mK and further increase in steps of 10~mK. 
The white cross marks the center of the SNR, and the white dashed line and solid line mark the inner and outer shells. 
The azimuthal sections (a, b c) for analysis are marked.
}
\label{images}
\end{center}
\end{figure*}

As the polarization gradient is insensitive to the change in the large-scale changes in polarization direction from Faraday rotation,
its power spectrum preserves the scaling slope of the magnetic turbulence.
We derived the magnitude of the polarization gradients $\nabla P$ at 5~GHz, according to~\citet{ghb11},

\begin{equation}
|\nabla \bm{P}| = \sqrt{\left(\frac{\partial Q}{\partial x}\right)^2 + \left(\frac{\partial U}{\partial x}\right)^2 + \left(\frac{\partial Q}{\partial y}\right)^2 + \left(\frac{\partial U}{\partial y}\right)^2} 
\end{equation}
where $x$ and $y$ are the horizontal and vertical axes of the image, and $\bm{P}=Q+iU$ is the complex polarization. 
As shown in Fig.~\ref{images}, the filamentary strips around the shell reveals the largest fluctuations of the magnetic field and gas density in the shock front. 

%
%

\section{Statistical analysis}  
\subsection{Structure function and Power spectrum}
Both the SF and the PS are common methods for characterizing random fields, and give the same scaling information on the energy cascade of MHD turbulence.

\begin{figure*}[!hbt]
\begin{center}
\includegraphics[angle=0,height=0.32\textwidth]{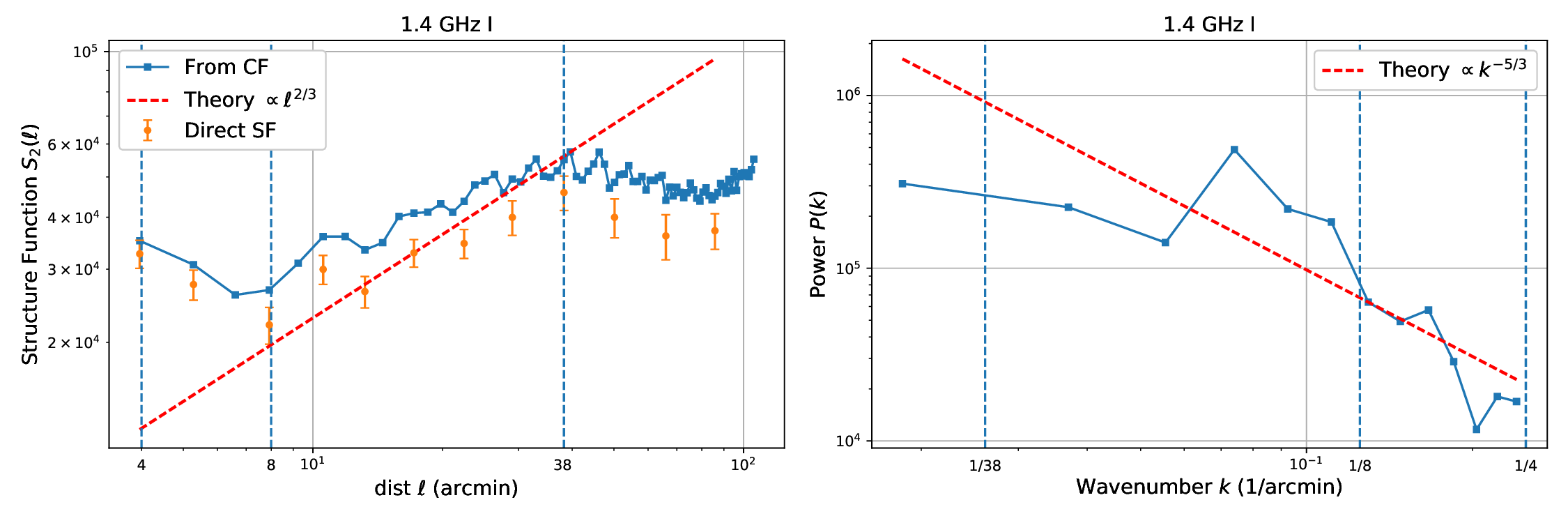}
\includegraphics[angle=0,height=0.32\textwidth]{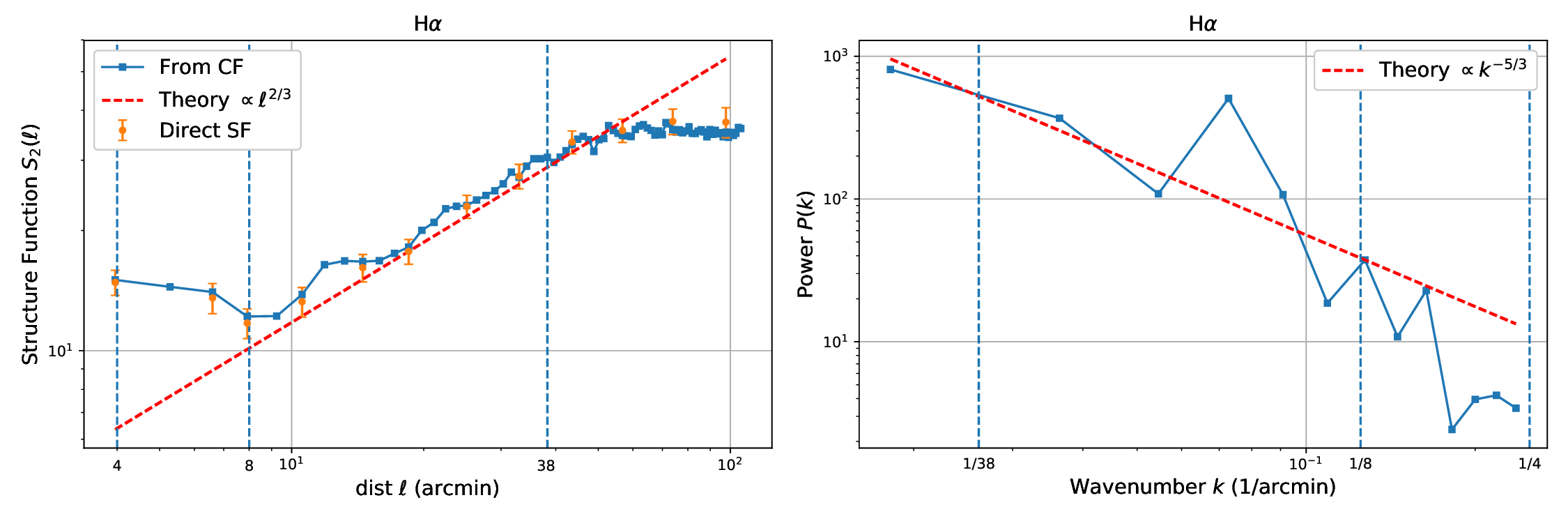}
\caption{ The 1D structure function of radio total intensity $I$ at 1.4~GHz (upper~panels) and the H$_{\alpha}$ images (lower panels) of the inner shell section of HB~9, and the
1D power spectrum derived from a FFT of the autocorrelation function.
The plots derived from the radio and H$_{\alpha}$ maps approximately represent the magnetic disturbances (upper) and the plasma density fluctuations (lower), respectively.
Vertical lines mark the scale of the angular resolution, the shell thickness and the averaged annulus radius of $l=$4$\arcmin$, $8 \arcmin$, and $38\arcmin$ (left), and corresponding wavenumbers (right).
The total length along the shell is 60$\arcmin$. The error bars for the structure function are also shown.
}
\label{21cmI_Ha}
\end{center}
\end{figure*}

For a given 2-dimensional (2D) physical quantity, the correlation function (CF) and the SF are defined as 
\begin{equation}\label{corrf}
CF(\bm{r})=\langle[f(\bm{x}+\bm{r})f(\bm{x})]\rangle ,
\end{equation}
\begin{equation}\label{sf}
SF(\bm{r})=\langle[f(\bm{x}+\bm{r})-f(\bm{x})]^2\rangle ,
\end{equation}
where $\bm{r}$ is a separation vector, and $\langle...\rangle$ denotes an averaging over the whole volume space.

The SF is related to the CF 
\begin{equation}\label{sf_cf}
SF(\bm{r})=2[CF(0)-CF(\bm{r})],
\end{equation}
According to the Wiener–Khinchin theorem, the PS can be obtained by performing a Fourier transform of the correlation function. 
\begin{equation}\label{ps}
P_{2D}(\bm{k}) = \frac{1}{(2\pi)^2} \int \langle f(\bm{x})f(\bm{x} + \bm{r}) \rangle e^{-i\bm{k}\cdot\bm{r}} d\bm{r}
\end{equation}
%
%

\begin{figure}[!hbt]
\begin{center}
\includegraphics[angle=0,height=0.32\textwidth]{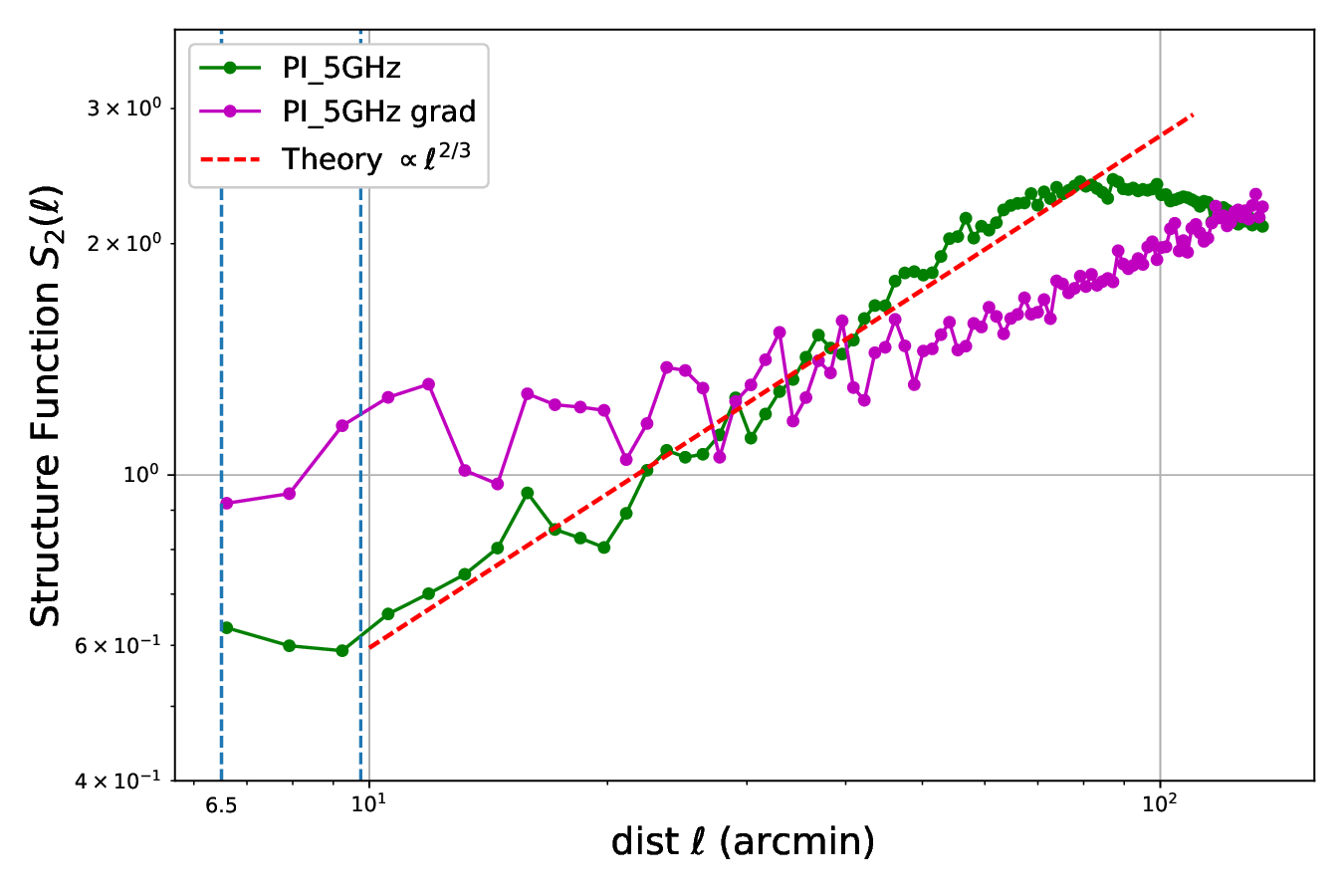}
\caption{The 1D structure function for $PI$ and $\nabla P$ data of the inner shell section of HB~9, calculated from the autocorrelation function.
Vertical lines mark the scale of the 5-GHz Nyquist limit and the angular resolution of $l=$6.5$\arcmin$ and $9.75 \arcmin$. 
}
\label{6cmSF}
\end{center}
\end{figure}

As MHD turbulence maintains a cascade power law of \(v \propto \delta B \propto k^{-m}\), where \(v\) is the velocity, $\delta B$ is the magnetic field distrubation, 
and $m=1/3$ stands for a Kolmogorov turbulence.
Then the derived PS has $E_{2D}(k)\propto k^{-2m-2}=k^{-8/3}$. 
The relationship holds for the 1D physical quantity, with a scalar distance $l$.
For 1D SF, the power-law is $l^{2/3}$, and the 1D power spectrum is $E_{1D}(k)\propto k^{-2m-1}=k^{-5/3}$.

\subsection{Application to data}
To investigate the turbulence properties in the inner shells (a as marked in Fig.~\ref{images}) of HB~9, we 
performed 1D SF and PS analyses firstly for the radio total intensity ($I$) and H$_{\alpha}$ emission. 
The center of the map ($\alpha_{2000}$, $\delta_{2000}$) =($5^{\rm{h}}01^{\rm{m}}0.0^{\rm{s}}$, +46$\degr$39$\arcmin$0.0$\arcsec$) is taken approximately as the center of the SNR. 
Given the nearly spherical shell morphology of HB~9, the inner shell was defined using an annular mask with inner and outer radii of 26 and 33~pixels,
based on the sharp radio emission boundary that marks the shock front. From this annulus, we extracted the azimuthal section between 195$\degr$ and 275$\degr$, 
which primarily covers the region of low polarization percentage at 5~GHz. 
We also selected two outer shell regions (b, c as marked in Fig.~\ref{images}) for comparision, with defined inner and outer radii of 39 and 45~pixels.
The first section covers the eastern outer shell (140$\degr - 260\degr$), where the 5-GHz polarization fraction is relatively high ($\sim 20-30$~\%). 
The second section covers the southern outer shell (270$\degr - 360\degr$), which exhibits a high RM value similar with the inner shell.
Within these angular range, we consider that the path length through the shell along the line‑of‑sight is approximately constant,
and thus geometric projection effects can be neglected.

We calculated the overbrightness map as $\delta I =I -\overline{I}$, where $\overline{I}$ is the mean brightness over all pixels within the shell region. 
The correlation function of the overbrightness images was then computed, from which we further derived the SF. The minimum separation distance is set to 1 pixel. 
The uncertainty is reflected by the variation among adjacent bins.
To cross‑check the results given the limited sample, we also computed the SF directly from data pairs, where the uncertainty was estimated from the standard deviation of the values within each distance bin. 
Additionally, we calculated the 1D PS by applying a Fourier transform to the correlation function. 

For the 5-GHz $PI$ and $\nabla P$, the resolution of 9.75$\arcmin$ is more than twice that of the FAST map.
The uncertainties are larger. We only present the 1D SF derived from the CF of HB~9.

To examine the turbulence behavior within the inner shell using an enlarged sample, we analyzed the 2D PS over the entire inner region of HB~9. 
Pixels outside the outer radius were masked and set to zero, to ensure that they do not contribute to the correlation. 
Then we obtained a radial power spectrum by azimuthally averaging over rings of constant wavenumber.

\begin{figure*}[!hbt]
\begin{center}
\includegraphics[angle=0,height=0.32\textwidth]{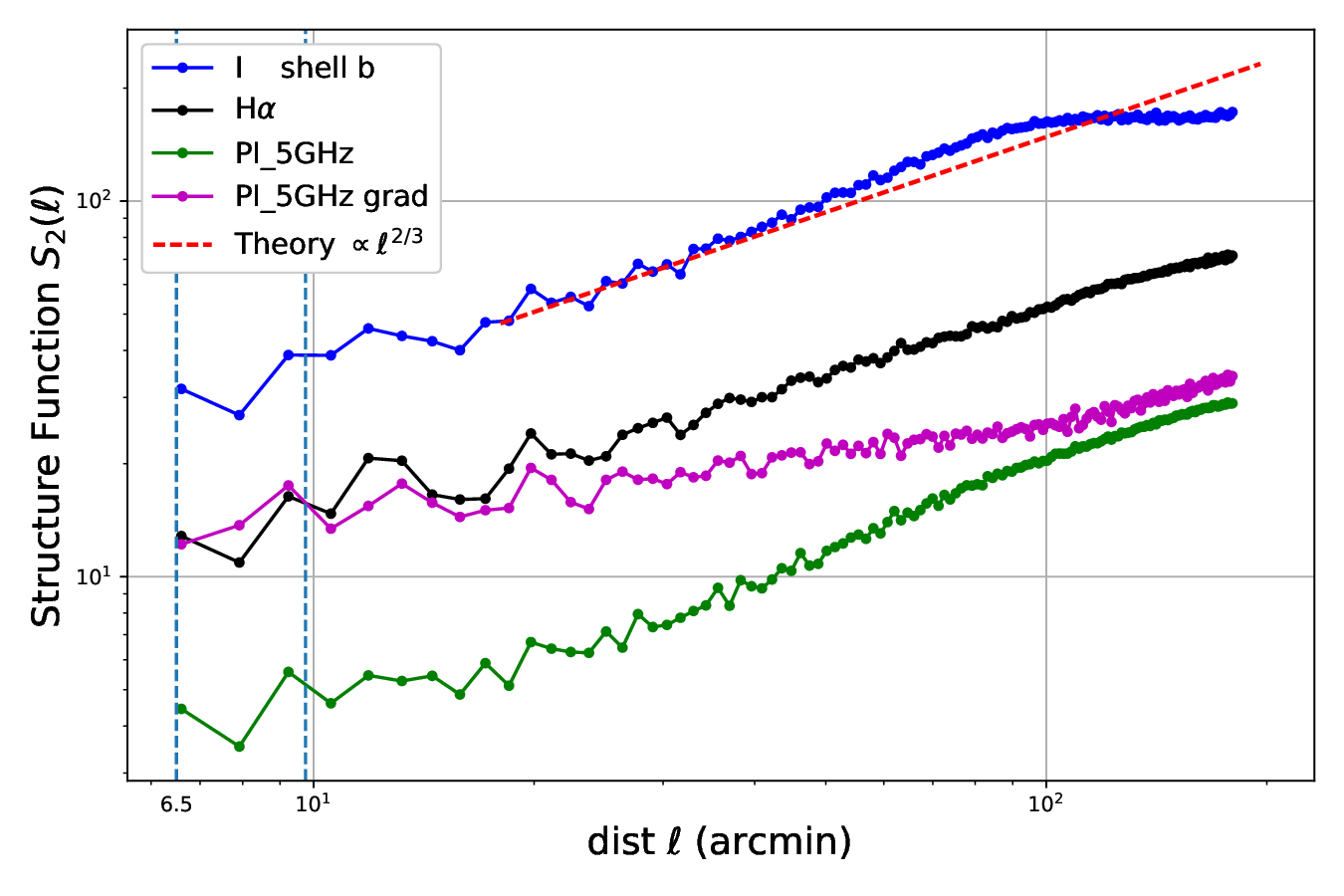}
\includegraphics[angle=0,height=0.32\textwidth]{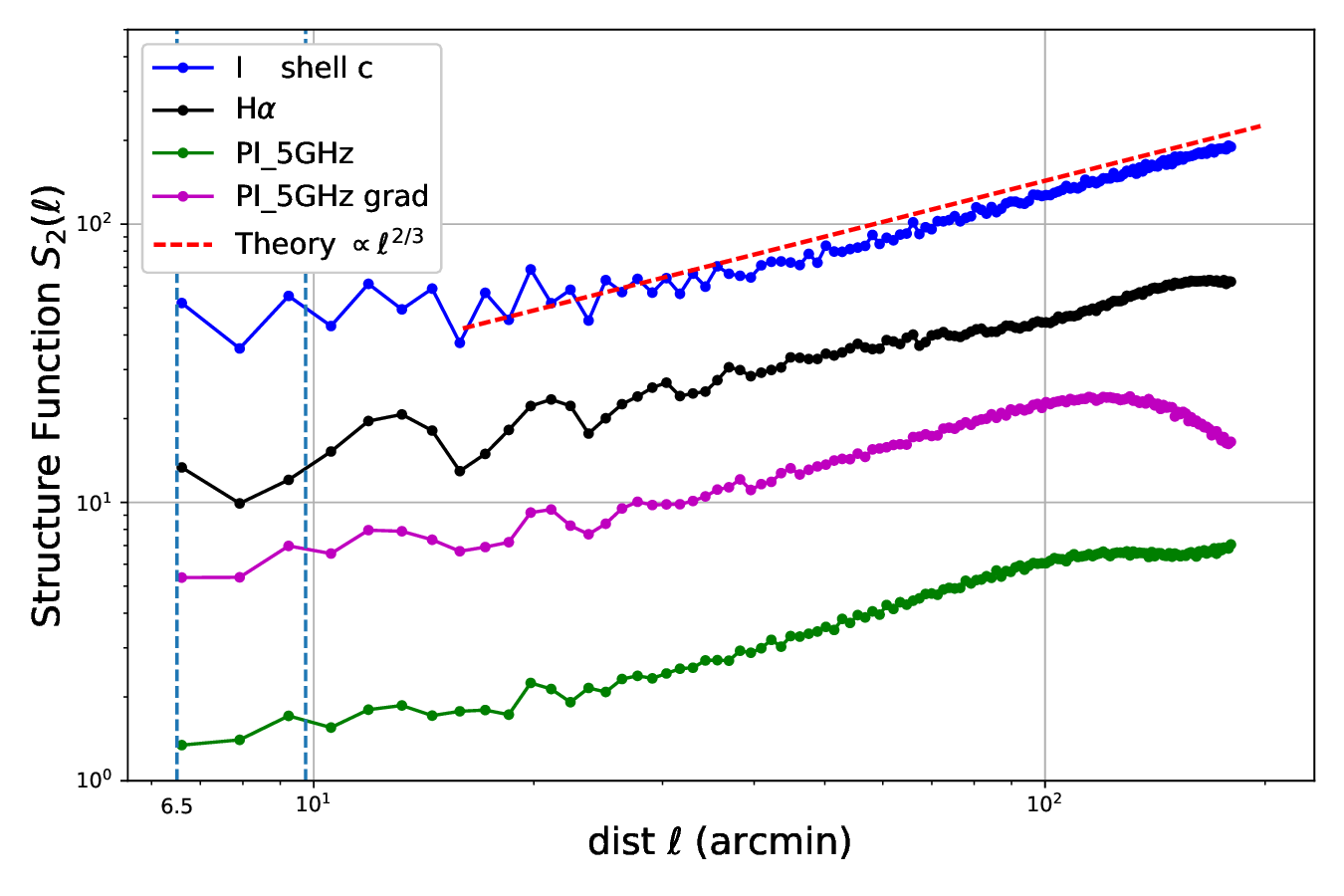}
\caption{The 1D structure function for 5-GHz $I$, $H_{\alpha}$, $PI$ and $\nabla P$ data of the outer shell sections (b, c) of HB~9, calculated from the autocorrelation function.
Vertical lines mark the scale of the 5-GHz Nyquist limit and the angular resolution of $l=$6.5$\arcmin$ and $9.75 \arcmin$.
}
\label{outshel}
\end{center}
\end{figure*}

\section{Results}
\subsection{1D SF and PS of the selected shell section}
Figure~\ref{21cmI_Ha} shows the 1D SF and PS of the radio total intensity ($I$) and H$_{\alpha}$ emission from the inner shell section of HB~9. 
The structure function was derived using two independent approaches: one from autocorrelation function, and the other from direct data pair calculation.
The results from the two methods are mutually consistent and follow a Kolmogorov scaling law of $l^{2/3}$.
This consistency indicates the presence of ``fully developed turbulence" in the inner shell section of HB~9, implying that the multiscale fluctuations are statistically stable in both time and space. 
From the SF for the 1.4-GHz radio total intensity, the scaling law holds within the range ($8\arcmin<l<38\arcmin$). 
At scales below 8$\arcmin$, the spectrum becomes flat. This deviation is likely attributable to the 7 pixels width of the annular region used in the analysis,
which may filter out the underlying spectral structures.
The 1D power spectra of both the radio total intensity ($I$) and H$_{\alpha}$ emission are broadly consistent with a Kolmogorov scaling slope of $-5/3$, although with
considerable variation possibly caused by the projection effect and the shell thickness including turbulence at different distances from the center.
For 1.4~GHz $I$, the SF calculated from the autocorrelation function shows an higher offset than that from the direct pair calculation, likely owing to a large-scale emission gradient under the limited sample.

Figure~\ref{6cmSF} shows the 1D SFs calculated from autocorrelation function for the 5-GHz $PI$ and $\nabla P$ in the inner shell section of HB~9. 
The spectral slope of $PI$ generally is consistent with the 1D Kolmogorov spectrum expected for fully developed turbulence with the range of ($10\arcmin<l<60\arcmin$).
The $\nabla P$ spectrum is shallower than the theoretical $l^{-2/3}$ law. 

Figure~\ref{outshel} presents the 1D SFs derived from the autocorrelatoin functions of $I$, H$_{\alpha}$, $PI$ and $\nabla P$ for the two outer shell regions (5~GHz $I$ is used to lower the amplitude).
Remarkably, both regions show a Kolmogorov-like scaling of $l^{2/3}$ over the inertial range, and it holds consistently across the selected sub-intervals.
This indicates that Kolmogorov-type MHD is also present in the outer shell. However, for the eastern outer shell, the SF of $\nabla P$ flattens
due to the dominance of well-ordered magnetic field, but $PI$ and $PA$ still exhibit sub-beam fluctuations.


\subsection{2D PS of the entire inner region}
\begin{figure}[!hbt]
\begin{center}
\includegraphics[angle=0,width=0.48\textwidth]{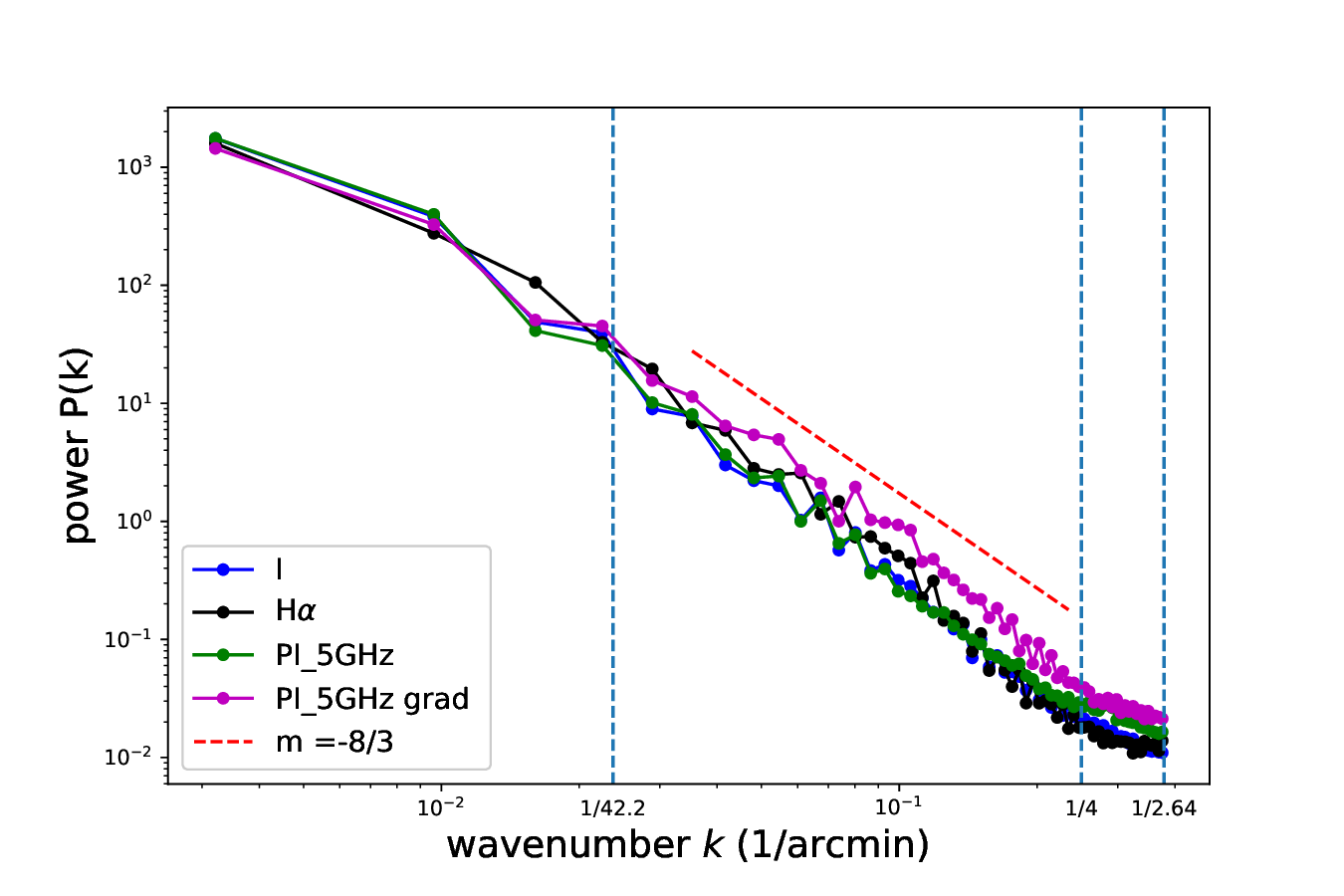}
\caption{The radial power spectra of the $I$, H$_{\alpha}$, $PI$, and $\nabla P$ data of the inner shell region of HB~9.
The red dashed line shows a power law with an index of $-8/3$.
The vertical lines mark the the wavenumnbers $k$ of $1/$2.64$\arcmin$, $1/4 \arcmin$ and the averaged annulus radius of $1/42.2\arcmin$.
}
\label{SF_all_inner}
\end{center}
\end{figure}

As shown in Figure~\ref{SF_all_inner}, the radially averaged power spectra of $I$, H$_{\alpha}$, $PI$ and $\nabla P$
derived from the 2D projected inner region of HB9, all exhibit a $k^{-8/3}$ power-law behavior in the inertial range, 
suggesting the presence of 3D Kolmogorov-type turbulence within the inner region, thus the add-up 2D-projection retains the same scaling.
At wavenumbers larger than $k=1/4\arcmin$, the spectra deviate upward from the scaling. This deviation is attributed to instrumental noise, 
whose flat power spectrum dominates on small scales. 
Notably, the Nyquist limit lies far above the expected dissipation scale of the turbulence (e.g., $\sim 10^{-6}R$ for Reynolds number $Re\sim 1$).
The power spectrum of the polarization gradient $\nabla P$ shows the power-law nature of magnetic turbulence,
may suggest a case of nearly isotropic turbulence in the inner shell~\citep{zl25}. 

\section{Discussion}
\subsection{H$\alpha$ and radio observations as diagnostic of the turbulence }
H$\alpha$ emission traces the fluctuations of electron density, which are passively mixed by the Alfvénic cascade in the MHD turbulence and therefore inherit the behavior
of the velocity field~\citep{gs95}. In contrast, radio synchrotron emission probes magnetic field fluctuations, which evolve on timescales distinct from the velocity cascade. 
Magnetic energy is generated at small scales via stretching and folding, and transfers to larger scales later. For fully developed turbulence (the velocity field has reached a
statistical steady state), the magnetic energy at large scales often does not have sufficient time to reach equipartition with the turbulent kinetic energy, 
resulting in a flat behavior at the large-scale end (Fig. 5 in~\citet{hxs22}. As shown in Fig.~\ref{21cmI_Ha}, the structure funtion of the radio emission for the inner shell of HB~9
begins to flatten at larger scales, while both the electron density and the magnetic field show a Kolmogorov-like scaling, indicating a fully developed turbulence.

\subsection{Turbulence mechanism in the inner shell}
The interaction between SNR shocks and the turbulent ISM has been intensively studied~\citep{gll12,iso13,ppm25}. 
In the early evolution of an SNR, the Richtmyer–Meshkov instability (RMI) plays a dominant role. Meanwhile, the Rayleigh–Taylor instability (RTI) grows in the contact discontinuity (CD)
region behind the reverse shock and can persist for a long time.

At age of HB~9 (4000-7000 yrs), the inner shell likely corresponds to the inner ejecta that have been crossed by the reverse shock. 
However, there are not any large-scale finger-like or mushroom-like structures, which would otherwise break the Kolmogorov spectrum. 
The absence of such features in the inner shell may be that they may 
have been smoothed out by turbulent mixing and overall expansion. 
~\citet{hxs22} simulated the interaction of a shock with an inhomogeneous medium and found that RMI-driven turbulence can maintain a Kolmogorov energy spectrum over a considerable timescale. 
Even when the preshock density distribution has a shallow power spectrum, vortex modes still dominate and evolve to follow Kolmogorov scaling. 
Furthermore, 2D MHD simulations by~\citet{ppm25} showed that when an SNR expands into a turbulent ISM with Kolmogorov-like spectra in both density and magnetic field, 
the turbulence inside the shell retains a Kolmogorov-type power spectrum up to an age of 3000 years, which is quite similar to the condition of the inner shell.

The observed constancy of RM across the inner shell (124~rad~m$^{-2}$ from~\citet{xzs25}) indicates that the line-of-sight magnetic field does not vary significantly on large scales.
A mean radial magnetic field is present, possibly generated by the stretching and folding action of the turbulent dynamo or by RMI itself.
Numerical simulations of RMI in SNR ejecta indeed show that the post-shock field can acquire a substantial radial component while remaining highly disordered on small scales~\citep{iso13,wjf17}.

\subsection{Different Random-to-Ordered Magnetic Field Ratios in the Inner and Outer Shells}
Despite both the inner and outer shells exhibiting Kolmogorov scaling, indicating the presence of fully developed MHD turbulence in both regions, 
the random-to-ordered magnetic field ratios are different. In the outer shell, the forward shock has compressed the interstellar materials,
generating a well-ordered, largely tangential magnetic field, as reflected by the a polarization fraction of $\sim 20-$30\% at 5~GHz. 

In contrast, in the inner shell, the RMI generates strong disordered magnetic fluctuations, and thereby amplifying the random field component.
We calculate the random-to-ordered magnetic field ratio $\delta B/\bar{B}$, from the fractional polarization ($\sim 5$\%) at 5~GHz, assuming neglectable other depolarization.
For a power-law CR electron energy index of $s= 2\alpha +1=2$, the intrinsic polarization fraction of a perfectly ordered field is about 70\%.  
We estimated the random component based on the analytic formulae (31) in ~\citet{bp16}, assuming that the magnetic field is a combination of an ordered component and 
a random component with isotropic Gaussian distribution. The ratio of the random component to the ordered component of the magnetic field $\delta B/\bar{B}$ is estimated as $\sim$2.4 for
the inner shell and $\sim$0.8 for the outer shell with a polarization fraction of 30\%. 
Adopting the ordered field $\bar{B}$ of 7.6~$\mu$G derived in~\citet{xzs25}, we derived an upper limit of $\delta B \sim 18~\mu$G for the inner shell.

~\citet{g25} summarized the magnetic field strengths of several young SNRs ($80-250~\mu$G) based on the widths of their X-ray synchrotron filaments. 
The ratio of the random component to the ordered component of the magnetic field $\delta B/\bar{B}$ was derived from the polarization degree measured by the Imaging X-ray Polarimetry Explorer (IXPE)
in~\citet{zpf23}, yielding values of 1.2, 1.8, and 3.3 for SN 1006, the Tycho rim, and Cas A, respectively. They discussed that the magnetic turbulence in SNRs is probably environment-dependent, with larger density fluctuations (e.g. knotty structures in Cas~A) leading to higher turbulence levels. The turbulence level in the inner shell of HB~9 is lower than that of Cas A but higher than that of Tycho,
possibly indicating that the inner shell contains clumpy structures with large density fluctuations too. Since HB~9 is an evolved SNR with an age of 
$4000-7000$~yr and a shock velocity of 100$-$200 km~s$^{-1}$ estimated from~\citet{llm24}, the inferred upper limit of $\delta B$ of 18~$\mu$G is reasonable.

\section{Summary}
We have conducted a systematic multi‑tracer statistical analysis of MHD turbulence for the inner and outer shells of SNR HB~9, using radio total intensity ($I$), H$_{\alpha}$ emission,
polarized intensity ($PI$), and polarization gradient ($\nabla P$) data. For the eastern inner shell section,
the 1D structure functions derived from $I$, H$_{\alpha}$, $PI$ all follow a power law $l^{2/3}$, and the corresponding 1D power spectra are consistent with $k^{-5/3}$. 
The 2D radial average power spectra for the full inner region all follow $k^{-8/3}$, including the polarization gradient ($\nabla P$). This consistency across independent observables provides strong evidence that the inner shell hosts fully developed, Kolmogorov‑type MHD turbulence.
We estimated the strength of random magnetic field therein from the low fractional polarization ($\sim$5\%), and obtained a upper limit of $\delta B \sim 18~\mu$G.
The observed Kolmogorov spectrum is consistent with RMI‑driven turbulence, as supported by MHD simulations of shock propagation through an homogeneous medium.

\begin{acknowledgements}
We thank the anonymous referee for constructive comments which helped to improve the paper. 
We thank Dr.~Yik Ki Ma for valuable comments and suggestions on the manuscript.
We acknowledge the support from the Key Program of National Natural Science Foundation of China (12433006), and the Guizhou Provincial Science and Technology Projects 
(No. QKHFQ[2023]003, No. QKHFQ[2024]001, No. QKHPTRC-ZDSYS[2023]003). This work made use of the data from Five-hundred-meter Aperture Spherical radio Telescope (FAST).
FAST is a Chinese national mega-science facility, operated by National Astronomical Observatories, Chinese Academy of Sciences.

\end{acknowledgements}

\bibliographystyle{aa}

\bibliography{G160.bib}

\begin{thebibliography}{30}
\expandafter\ifx\csname natexlab\endcsname\relax\def\natexlab#1{#1}\fi

\bibitem[{{Bandiera} \& {Petruk}(2016)}]{bp16}
{Bandiera}, R. \& {Petruk}, O. 2016, \mnras, 459, 178

\bibitem[{{Beresnyak}(2019)}]{b19}
{Beresnyak}, A. 2019, Living Reviews in Computational Astrophysics, 5, 2

\bibitem[{{Blandford} \& {Ostriker}(1978)}]{bo78}
{Blandford}, R.~D. \& {Ostriker}, J.~P. 1978, \apjl, 221, L29

\bibitem[{{Cho} \& {Lazarian}(2002)}]{cl02}
{Cho}, J. \& {Lazarian}, A. 2002, \prl, 88, 245001

\bibitem[{{Gaensler} {et~al.}(2011){Gaensler}, {Haverkorn}, {Burkhart},
  {Newton-McGee}, {Ekers}, {Lazarian}, {McClure-Griffiths}, {Robishaw},
  {Dickey}, \& {Green}}]{ghb11}
{Gaensler}, B.~M., {Haverkorn}, M., {Burkhart}, B., {et~al.} 2011, \nat, 478,
  214

\bibitem[{{Gao} {et~al.}(2011){Gao}, {Han}, {Reich}, {Reich}, {Sun}, \&
  {Xiao}}]{ghr11}
{Gao}, X.~Y., {Han}, J.~L., {Reich}, W., {et~al.} 2011, \aap, 529, A159

\bibitem[{{Goldreich} \& {Sridhar}(1995)}]{gs95}
{Goldreich}, P. \& {Sridhar}, S. 1995, \apj, 438, 763

\bibitem[{{Greco}(2025)}]{g25}
{Greco}, E. 2025, Frontiers in Astronomy and Space Sciences, 12, 1717808

\bibitem[{{Guo} {et~al.}(2012){Guo}, {Li}, {Li}, {Giacalone}, {Jokipii}, \&
  {Li}}]{gll12}
{Guo}, F., {Li}, S., {Li}, H., {et~al.} 2012, \apj, 747, 98

\bibitem[{{Haffner} {et~al.}(1998){Haffner}, {Reynolds}, \& {Tufte}}]{hrt98}
{Haffner}, L.~M., {Reynolds}, R.~J., \& {Tufte}, S.~L. 1998, \apjl, 501, L83

\bibitem[{{Hu} {et~al.}(2022){Hu}, {Xu}, {Stone}, \& {Lazarian}}]{hxs22}
{Hu}, Y., {Xu}, S., {Stone}, J.~M., \& {Lazarian}, A. 2022, \apj, 941, 133

\bibitem[{{Inoue} {et~al.}(2013){Inoue}, {Shimoda}, {Ohira}, \&
  {Yamazaki}}]{iso13}
{Inoue}, T., {Shimoda}, J., {Ohira}, Y., \& {Yamazaki}, R. 2013, \apjl, 772,
  L20

\bibitem[{{Jiang} {et~al.}(2019){Jiang}, {Yue}, {Gan}, {Yao}, {Li}, {Pan},
  {Sun}, {Yu}, {Liu}, {Tang}, {Qian}, {Lu}, {Yan}, {Peng}, {Zhang}, {Wang},
  {Li}, \& {Li}}]{jyg19}
{Jiang}, P., {Yue}, Y., {Gan}, H., {et~al.} 2019, Science China Physics,
  Mechanics, and Astronomy, 62, 959502

\bibitem[{{Lazarian} \& {Pogosyan}(2012)}]{lp12}
{Lazarian}, A. \& {Pogosyan}, D. 2012, \apj, 747, 5

\bibitem[{{Lazarian} \& {Pogosyan}(2016)}]{lp16}
{Lazarian}, A. \& {Pogosyan}, D. 2016, \apj, 818, 178

\bibitem[{{Leahy} {et~al.}(2020){Leahy}, {Ranasinghe}, \& {Gelowitz}}]{lrg20}
{Leahy}, D.~A., {Ranasinghe}, S., \& {Gelowitz}, M. 2020, \apjs, 248, 16

\bibitem[{{Li} {et~al.}(2024){Li}, {Lu}, {Mao}, {Xia}, {Chen}, {Zhou}, \&
  {Zhou}}]{llm24}
{Li}, J.-T., {Lu}, L.-Y., {Mao}, H., {et~al.} 2024, \aap, 690, A42

\bibitem[{{Petruk} \& {Kuzyo}(2025)}]{pk25}
{Petruk}, O. \& {Kuzyo}, T. 2025, \apj, 994, 147

\bibitem[{{Prete} {et~al.}(2025){Prete}, {Perri}, {Meringolo}, {Primavera}, \&
  {Servidio}}]{ppm25}
{Prete}, G., {Perri}, S., {Meringolo}, C., {Primavera}, L., \& {Servidio}, S.
  2025, \apjs, 277, 44

\bibitem[{{Saha} {et~al.}(2021){Saha}, {Bharadwaj}, {Chakravorty}, {Roy},
  {Choudhuri}, {G{\"u}nther}, \& {Smith}}]{sbc21}
{Saha}, P., {Bharadwaj}, S., {Chakravorty}, S., {et~al.} 2021, \mnras, 502,
  5313

\bibitem[{{Saha} {et~al.}(2019){Saha}, {Bharadwaj}, {Roy}, {Choudhuri}, \&
  {Chattopadhyay}}]{sbr19}
{Saha}, P., {Bharadwaj}, S., {Roy}, N., {Choudhuri}, S., \& {Chattopadhyay}, D.
  2019, \mnras, 489, 5866

\bibitem[{{Sezer} {et~al.}(2019){Sezer}, {Ergin}, {Yamazaki}, {Sano}, \&
  {Fukui}}]{sey19}
{Sezer}, A., {Ergin}, T., {Yamazaki}, R., {Sano}, H., \& {Fukui}, Y. 2019,
  \mnras, 489, 4300

\bibitem[{{Shanahan} {et~al.}(2023){Shanahan}, {Stil}, {Anderson}, {Beuther},
  {Goldsmith}, {Klessen}, {Rugel}, \& {Soler}}]{ssa23}
{Shanahan}, R., {Stil}, J.~M., {Anderson}, L., {et~al.} 2023, \apj, 957, 60

\bibitem[{{Shimoda} {et~al.}(2018){Shimoda}, {Akahori}, {Lazarian}, {Inoue}, \&
  {Fujita}}]{sal18}
{Shimoda}, J., {Akahori}, T., {Lazarian}, A., {Inoue}, T., \& {Fujita}, Y.
  2018, \mnras, 480, 2200

\bibitem[{{West} {et~al.}(2017){West}, {Jaffe}, {Ferrand}, {Safi-Harb}, \&
  {Gaensler}}]{wjf17}
{West}, J.~L., {Jaffe}, T., {Ferrand}, G., {Safi-Harb}, S., \& {Gaensler},
  B.~M. 2017, \apjl, 849, L22

\bibitem[{{Xiao} {et~al.}(2025){Xiao}, {Zhu}, {Sun}, {Reich}, {Reich}, {Jiang},
  \& {Sun}}]{xzs25}
{Xiao}, L., {Zhu}, M., {Sun}, X.-H., {et~al.} 2025, \aap, 697, A131

\bibitem[{{Zhang} \& {Liu}(2025)}]{zl25}
{Zhang}, J.-F. \& {Liu}, Z.-Q. 2025, \apj, 985, 235

\bibitem[{{Zhao} {et~al.}(2020){Zhao}, {Jiang}, {Li}, {Chen}, {Yu}, \&
  {Wang}}]{zjl20}
{Zhao}, H., {Jiang}, B., {Li}, J., {et~al.} 2020, \apj, 891, 137

\bibitem[{{Zhou} {et~al.}(2023){Zhou}, {Prokhorov}, {Ferrazzoli}, {Yang},
  {Slane}, {Vink}, {Silvestri}, {Bucciantini}, {Reynoso}, {Moffett},
  {Soffitta}, {Swartz}, {Kaaret}, {Baldini}, {Costa}, {Ng}, {Kim},
  {Doroshenko}, {Ehlert}, {Heyl}, {Marin}, {Mizuno}, {Pesce-Rollins},
  {Sgr{\`o}}, {Tamagawa}, {Weisskopf}, {Xie}, {Agudo}, {Antonelli}, {Bachetti},
  {Baumgartner}, {Bellazzini}, {Bianchi}, {Bongiorno}, {Bonino}, {Brez},
  {Capitanio}, {Castellano}, {Cavazzuti}, {Chen}, {Ciprini}, {De Rosa}, {Del
  Monte}, {Di Gesu}, {Di Lalla}, {Di Marco}, {Donnarumma}, {Dov{\v{c}}iak},
  {Enoto}, {Evangelista}, {Fabiani}, {Garcia}, {Gunji}, {Hayashida}, {Iwakiri},
  {Jorstad}, {Kislat}, {Karas}, {Kitaguchi}, {Kolodziejczak}, {Krawczynski},
  {La Monaca}, {Latronico}, {Liodakis}, {Maldera}, {Manfreda}, {Marinucci},
  {Marscher}, {Marshall}, {Matt}, {Mitsuishi}, {Muleri}, {Negro}, {O'Dell},
  {Omodei}, {Oppedisano}, {Papitto}, {Pavlov}, {Peirson}, {Perri}, {Petrucci},
  {Pilia}, {Possenti}, {Poutanen}, {Puccetti}, {Ramsey}, {Rankin}, {Ratheesh},
  {Roberts}, {Romani}, {Spandre}, {Tavecchio}, {Taverna}, {Tawara}, {Tennant},
  {Thomas}, {Tombesi}, {Trois}, {Tsygankov}, {Turolla}, {Wu}, \&
  {Zane}}]{zpf23}
{Zhou}, P., {Prokhorov}, D., {Ferrazzoli}, R., {et~al.} 2023, \apj, 957, 55

\bibitem[{{Ziegenbalg}(2025)}]{z25}
{Ziegenbalg}, S. 2025, Research Notes of the American Astronomical Society, 9,
  227

\end{thebibliography}

\end{document}